\documentstyle[epsf,twocolumn]{jpsj}
\author{Yusuke {\sc Takazawa},Yoshiki {\sc Imai} and Norio {\sc Kawakami}}
\title{
Electron Transport through T-Shaped Double-Dots System}
\inst{Department of Applied Physics, 
Osaka University, Suita, Osaka 565-0871}

\recdate{}

\abst{Correlation effects on 
electron transport through a system of T-shaped double-dots 
are investigated, for which only one of the dots is directly 
connected to the leads.  We evaluate the local density of states
and the conductance by means of  the 
non-crossing approximation
at finite temperatures as well as the slave-boson 
mean field approximation at zero temperature.
It is found that the dot which 
is not directly connected to the leads 
considerably influences  the 
conductance, making its behavior quite  different 
from the case of a single-dot system.
In particular, we find a novel phenomenon 
 in the Kondo regime with a small inter-dot coupling, i.e. 
 Fano-like suppression of the 
Kondo-mediated conductance, when 
two dot levels coincide with each other energetically.
}

\kword{Kondo effect, interference effect, double dots }

\def\runtitle{Electron Transport through T-Shaped Double-Dots System} 
\def\runauthor{Yusuke {\sc Takazawa},
Yoshiki {\sc Imai} and Norio {\sc Kawakami}}

\begin{document}
\sloppy
\maketitle

\section{Introduction}
Recent intensive investigations of electron
 transport phenomena through quantum dot systems
have revealed  that the strong electron-electron interaction
plays a crucial role at low temperatures. In particular,
the Kondo effect, which is a prototypical 
many-body effect, has been discovered in a single-dot system,
stimulating further theoretical and  experimental
studies\cite{glazman,meir1,wingreen,kawabata,oguri,gold,cronenwwett,schmid,simmel,tarucha}. The Kondo effect has been also observed in 
coupled quantum dots\cite{austing,waugh,blick,Oosterkamp}, in which the inter-dot coupling is 
coulombly controllable, making this system 
especially fascinating. In particular, from systematic 
studies on coupled quantum dots it has been pointed out that
a wide variety of Kondo-like  phenomena can indeed occur
at low temperatures\cite{aono1,aono2,sakai1,izumida,hofstetter,vojta}.

There has been another remarkable phenomenon observed 
in quantum dot systems, i.e. the Fano effect\cite{fano,bogdan,hofstetter2,Gores}, which is 
caused by the interference between two distinct
resonant channels in the tunneling process.  
When this effect is combined with the Kondo effect, it may 
provide various  interesting many-body effects by the interplay of 
 both phenomena.

\begin{figure}[h]
\epsfxsize=11cm
\centerline{\epsfbox{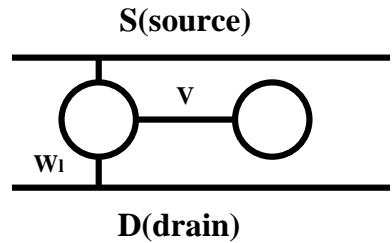}}
\vspace{-35mm}
\caption{T-shaped double-dots system in parallel geometry.
The left dot is
connected to both of the leads 
whereas the right dot is  to the left one. 
The tunneling amplitude  between the dots is $V$ and that
between  the  left dot and the leads is $W_{l}$.
}
\label{fig:dqd}
\end{figure}

In this paper, we investigate tunneling phenomena through a system 
of two coupled quantum dots in  specific parallel-geometry 
shown in Fig.\ref{fig:dqd}, which 
may be referred to as a system of T-shaped double dots.
 The system is also regarded as 
a single-dot system decollated by another dot,
so that the additional right dot may open a 
tunneling channel distinct from the direct resonance via the left dot,
 and is thus expected to give rise to the interplay between the 
Kondo effect and the Fano-type interference effect. 
By using non-crossing approximation (NCA)
\cite{coleman,kuramoto,bickers,puruschke,imai} 
approach for a 
generalized Anderson tunneling Hamiltonian, we calculate 
the local density of states (DOS) and the tunneling conductance. 
The NCA approach is further supplemented  by the zero-temperature 
calculation by means of slave-boson mean field 
(SBMF) approximation\cite{jones,hewson}.
We discuss how several tunable parameters influence the local 
DOS and the tunneling conductance. It is shown
that the inter-dot coupling and the relative position of
dot levels play an essential role for  electron transport. 
In particular, we find that in the Kondo regime with a small 
inter-dot tunneling, the Kondo-mediated conductance 
is suppressed by the Fano-like interference effect.

This paper is organized as follows. In the next section, we briefly 
describe the model and the method, and then in  \S 3 we show the results 
obtained for the local DOS and the conductance. 
A brief summary is given in  \S 4

\section{Model and Method}

\subsection{Model Hamiltonian}
We consider T-shaped double dots 
shown in Fig. \ref{fig:dqd}. This system is described by 
an extended  Anderson Hamiltonian for double dots,
\begin{eqnarray}
&H&=H_{loc}+ H_{mix}\nonumber \\
&H&_{loc}=\sum_{i=l,r}\sum_{\sigma}\epsilon_{i}
d_{i\sigma}^{\dagger}d_{i\sigma}
+V\sum_{i\neq j}d_{i\sigma}^{\dagger}d_{i\sigma}
+U\sum_{i}n_{i\uparrow}n_{i\downarrow}\nonumber \\
&H&_{med}=\sum_{k\sigma \alpha}\epsilon_{k\sigma \alpha}
c_{k\sigma \alpha}^{\dagger}c_{k\sigma \alpha}
+\frac{1}{\sqrt{2}}\sum_{k\sigma \alpha}(W_{l}
c_{k\sigma \alpha}^{\dagger}d_{l\sigma}+h.c)
\label{eqn:H}
\end{eqnarray}
where $d_{i\sigma }^{\dagger}$ creates an electron in 
the left or right dot (labeled by $i=l,r$) with  the
 energy $\epsilon_{i}$ and  spin $\sigma$, whereas
 $c_{k\sigma \alpha}^{\dagger}$ creates an electron in 
the leads specified by $\alpha=s,d$ (source, drain) with the
energy $\epsilon_{k\sigma \alpha}$ and spin $\sigma $. 
$V$ is the inter-dot tunneling amplitude 
while $W_{l}$ is that
between the left dot and the leads. We assume that the intra-dot Coulomb 
interaction is sufficiently large to prohibit the double 
occupancy of electrons in each dot.  To treat this situation, 
we apply the NCA to our model, which is known to 
give reasonable results in the temperature range 
larger than the Kondo temperature.\cite{coleman,kuramoto,bickers}
At lower temperatures
the  NCA may become inappropriate so that 
we adopt a complementary approach  based on the 
SBMF approximation at zero temperature.\cite{jones,hewson}

\subsection{NCA approach}

Let us treat the Hamiltonian  (\ref{eqn:H})
by extending the NCA to the case possessing
 several kinds of ionic propagators. A similar extension of the 
NCA to the case of four ionic propagators 
has been applied successfully to the single impurity
 Anderson model with the finite Coulomb
interaction\cite{puruschke}. In order to formulate the Green function 
for our double-dots model,
 we first diagonalize the local 
Hamiltonian $H_{loc}$,
which  can be expressed in terms 
of the Hubbard operators $X_{mn}=|m\rangle \langle n|$ as
\begin{eqnarray}
H_{loc}=\sum_{m=1}^{9}E_{m}X_{mm},
\end{eqnarray}
where $|m\rangle $ and $E_{m}$ respectively 
 denote the eigenstate of the local Hamiltonian  and 
the corresponding eigenenergy. Recall that we should impose the 
constraint on the Hilbert space
such that double occupancy of electrons in each dot is 
forbidden. The remaining 
Hamiltonian  $H_{med}$ can be also expressed in terms 
of the Hubbard operators as
\begin{eqnarray}
H_{med}&=&\sum_{k\sigma \alpha}\epsilon_{k\sigma \alpha}c_{k\sigma \alpha}^{\dagger}c_{k\sigma \alpha}\nonumber \\
&+&
\frac{1}{\sqrt{2}}\sum_{k\sigma \alpha}\sum_{mn}(W_{l}U_{nm}^{l\sigma * }c_{k\sigma \alpha}^{\dagger}X_{nm}+h.c),
\end{eqnarray}
where the matrix element $U_{nm}^{l\sigma }$ is given by
\begin{eqnarray}
U_{nm}^{l\sigma }=\langle m|d_{l\sigma }^{\dagger }|n\rangle.
\end{eqnarray}
For each eigenstate $|m\rangle $, the ionic propagator 
and the corresponding local DOS are defined by
\begin{eqnarray}
P_{m}(\omega )=\frac{1}{\omega -E_{m}-\Sigma_{m}(\omega )},
\end{eqnarray}
\begin{eqnarray}
\rho_{m}(\omega )=-\frac{1}{\pi }{\rm Im}P_{m}(\omega +{\rm i}\delta),
\end{eqnarray}
and its self-energy is represented in terms 
of $U_{nm}^{l\sigma }$ as
\begin{eqnarray}
\Sigma_{m}(\omega )&=&\sum_{n\sigma }
(|U_{mn}^{l\sigma }|^{2}+|U_{nm}^{l\sigma }|^{2})\nonumber \\
&\times&
\Delta \int_{-\infty }^{\infty }{\rm d}
\epsilon f(\eta_{mn} \epsilon )P_{n}(\omega +\eta_{mn} \epsilon )
\end{eqnarray}
where 
 $f(\epsilon)=[\exp (\beta \epsilon )+1]^{-1}$
 is  the Fermi distribution function ($\beta=1/T$), and
we set $\eta_{mn}=-1(+1)$ if the number of particles
accommodated in $|m\rangle $ is larger (smaller) than that
 in $|n\rangle $. 
We evaluate the self-energy $\Sigma_{m}(\omega )$ in 
a self-consistent perturbation theory up to the
second order in the hybridization $W_{l}$ as
is usually done  in the NCA. 
The left and right Green's functions are given by 
\begin{eqnarray}
G_{l(r)}(\omega )=\sum_{mn}|U_{mn}^{l(r)\sigma }|^{2}\langle 
\langle X_{mn};X_{nm}\rangle \rangle_{\omega },
\end{eqnarray}
\begin{eqnarray}
\langle \langle X_{mn}&;&X_{nm}\rangle \rangle_{\omega }\nonumber \\
&=&\frac{1}{Z_{loc}}\int {\rm d}\epsilon 
e^{-\beta \epsilon}[\rho_{m}(\epsilon )P_{n}
(\epsilon +\omega )-\rho_{n}(\epsilon )P_{m}(\epsilon -\omega )],
\end{eqnarray}
\begin{eqnarray}
Z_{loc}=\sum_{m=1}^{9}\int {\rm d}\epsilon 
e^{-\beta \epsilon }\rho_{m}(\epsilon ).
\end{eqnarray}
The local 
DOS for the $i$-th dot is thus expressed as
\begin{eqnarray}
\rho_{i}(\omega )=-\frac{1}{\pi }{\rm Im}G_{i}(\omega +{\rm i}\delta),
\end{eqnarray}
and the conductance is
\begin{eqnarray}
G=\frac{2e^{2}}{h}\sum_{\sigma}\Gamma\int_{-\infty}^{\infty}
{\rm d}\omega \left(-\frac{\partial f(\epsilon)}{\partial 
\epsilon}\right)\rho_{l}(\omega),
\end{eqnarray}
where $\Gamma$ is the tunneling strength,
which is given by the source (drain) tunneling strength 
$\Gamma^{s}$ ($\Gamma^{d}$) via the relation
$\Gamma=\frac{\Gamma^{s}\Gamma^{d}}{\Gamma^{s}+\Gamma^{d}}$.
We numerically iterate the procedure outlined above
to obtain the conductance  by extending the way
of Y.Meir ${\it et}$ ${\it al}$ \cite{meir2}for the single-dot system. 
In the following discussions, we assume that the hybridization 
between the left dot and the leads  is given by
$\Gamma^{s}=\Gamma^{d}=2\Delta$ for simplicity.

Here, we make a brief comment on the 
vertex correction in our NCA formalism.
It is known for the finite-$U$ Anderson model that
the vertex  corrections  in the local
Green function are not so important as far as  the one-particle 
spectral function is concerned.\cite{puruschke}
 It is expected that
this is also the case for the present system,
so that we neglect them in the following discussions. 

\subsection{ Slave-boson mean-field approach}
Since the NCA approach may not give reliable results 
at lower temperatures,
we  perform complementary calculations of the conductance 
at zero temperature by means  of the SBMF theory.\cite{jones,hewson}
We introduce a slave boson operator $b_{i}^{\dagger }$ and a 
fermion operator $f_{i\sigma }^{\dagger }$, where $b_{i}^{\dagger }$
($f_{i\sigma }^{\dagger }$) creates an empty state (a singly 
occupied state) in each dot under the constraint,
\begin{eqnarray}
\sum_{\sigma }f_{i\sigma }^{\dagger }f_{i\sigma }
+b_{i}^{\dagger }b_{i}=1.
\end{eqnarray}
In this method, the mixing term such as 
$W_{l}c_{k\sigma \alpha}^{\dagger}d_{l\sigma}$ in (\ref{eqn:H})
is replaced 
by $W_{l}c_{k\sigma \alpha}^{\dagger}f_{l\sigma}b_l^\dagger$,
etc. We now apply a mean-field treatment to our model by 
replacing the boson operator by its mean value; $b_{i}, b_{i}^{\dagger } 
\rightarrow \bar b_{i}$. The Hamiltonian is 
then written as 
\begin{eqnarray}
H&=&\sum_{i=l,r}\sum_{\sigma}\tilde{\epsilon_{i}}
f_{i\sigma }^{\dagger }f_{i\sigma }
+\tilde{V}\sum_{i\neq j}f_{i\sigma }^{\dagger }
f_{j\sigma }
+ \sum_{i}\lambda_{i}(\bar b_{i}^{2}-1)
\nonumber \\
&+&
\sum_{k\sigma \alpha}\epsilon_{k\sigma \alpha}
c_{k\sigma \alpha}^{\dagger}c_{k\sigma \alpha}
+ \sum_{k\sigma \alpha}(\tilde{W_{l}}
c_{k\sigma \alpha}^{\dagger}f_{l\sigma }+h.c)
\label{eqn:slv}
\end{eqnarray}
where we have introduced the Lagrange multipliers $\lambda_{i}$, 
the renormalized energy level 
in the $i$-th dot $\tilde{\epsilon_{i}}=\epsilon_{i}+\lambda_{i}$ 
the renormalized tunneling $\tilde{V}=V \bar b_{l} \bar b_{r}$, and 
$\tilde{W_{l}}=\frac{1}{\sqrt{2}}W_{l} \bar b_{l}$. For the 
mean-field Hamiltonian (\ref{eqn:slv}), we calculate the 
ground-state energy $E_{gs}$ as
\begin{eqnarray}
Egs&=&\frac{2}{\pi }{\rm Im}\sum_{P=\pm }[\xi_{P}
\ln (-\xi_{P})+(D+\xi_{P})\ln(-D-\xi_{P})]\nonumber \\
&+&
\sum_{i}\lambda_{i}(\bar b_{i}^{2}-1)+\beta (D+\mu )\pi,
\end{eqnarray}
where $\xi_{\pm}=\frac{1}{2}[\tilde{\epsilon_{r}}+
\tilde{\epsilon_{l}}-i\tilde{\Delta}
\pm\sqrt{(\tilde{\epsilon_{r}}-\tilde{\epsilon_{l}}+i\tilde{\Delta})^{2}
+4\tilde{v}^{2}}]$. 
The mean-field values of  $\lambda_{i}$ and $\bar b_{i}$ can be 
determined self-consistently by minimizing
$E_{gs}$ with respect 
to $\lambda_{i}$ and $b_{i}$, or equivalently  
with respect to the renormalized parameters $\tilde{\epsilon_{i}}$,
 $\tilde{V}$ and $\tilde{W_{l}}$. Then we can determine 
the conductance $G$ by calculating the Green's functions 
for both dots. The conductance is given by
\begin{eqnarray}
G=\frac{2e^{2}}{h}\sum_{\sigma}T(\omega=0),
\end{eqnarray}
where $T(\omega=0)=\Gamma\rho_{l}(\omega=0)$ is the transmission 
probability.

This completes the formulation based on the 
NCA  and the SBMF treatment.
By  calculating  the  local DOS
 in each dot from the above formulae, we can evaluate the 
tunneling conductance.

\section{ Numerical Results }

We now discuss how transport properties of  the T-shaped 
double-dots  system
are affected by the interference of two dots.  
We show the results obtained by the NCA at finite temperatures 
together with the complementary SBMF results
at zero temperature. We will use $\Gamma$ as 
the unit of the energy.

\subsection{ Tuning the inter-dot coupling}

\begin{figure}[h]
\epsfxsize=9cm
\centerline{\epsfbox{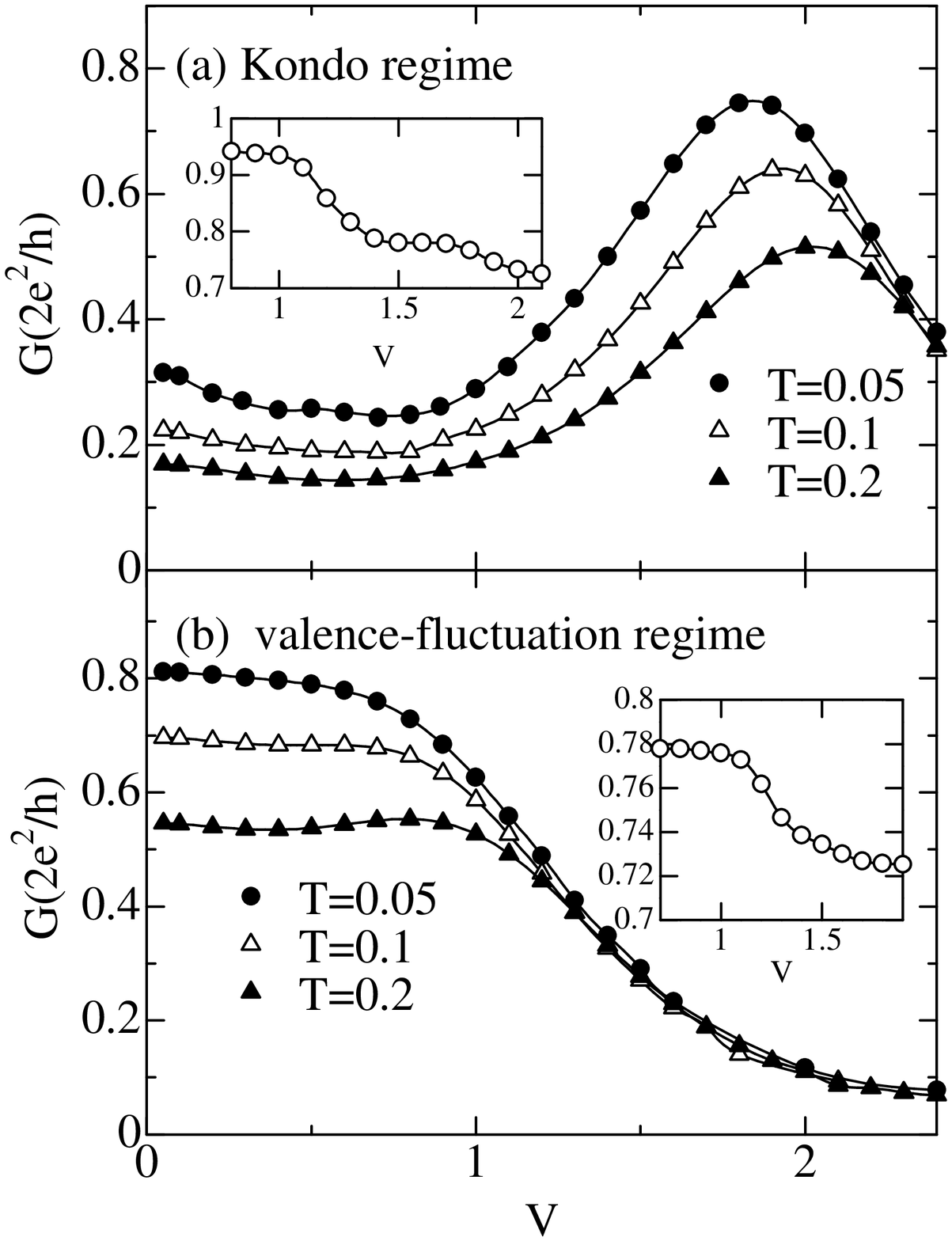}}
\vspace{-20mm}
\caption{Conductance for the T-shaped double-dots system as a 
function of the inter-dot coupling $V$ at various 
temperatures: (a) Kondo regime with the bare energy 
levels  $\epsilon_l=-2.0$ and $\epsilon_r=-3.0$,
(b) valence-fluctuation regime with
$\epsilon_l=-1.0$ and $\epsilon_r=-2.0$.
The inset shows the conductance computed by the SBMF
approximation at zero temperature. In the Kondo regime (a),
the electron number occupying the left (right) dot
is $n_l\simeq 1.0$ ($n_r\simeq 1.0$) for $V=0.1$, and 
$n_l\simeq 1.0$ ($n_r\simeq 0.9$) for $V=1.5$. 
On the other hand, in the valence fluctuation regime (b),
$n_l\simeq 0.6$ ($n_r\simeq 1.0$) for $V=0.1$, and 
$n_l\simeq 0.5$ ($n_r\simeq 0.5$) for $V=1.5$. 
}
\label{fig:Gvdepend}
\end{figure}

We first discuss the results 
by changing the inter-dot coupling $V$ systematically.
The calculated  conductance
is plotted in Fig. \ref{fig:Gvdepend}
by choosing typical sets of the parameters
 corresponding roughly to (a) the Kondo regime and (b) the 
 valence fluctuation regime.
 
\begin{figure}[h]
\epsfxsize=9cm
\centerline{\epsfbox{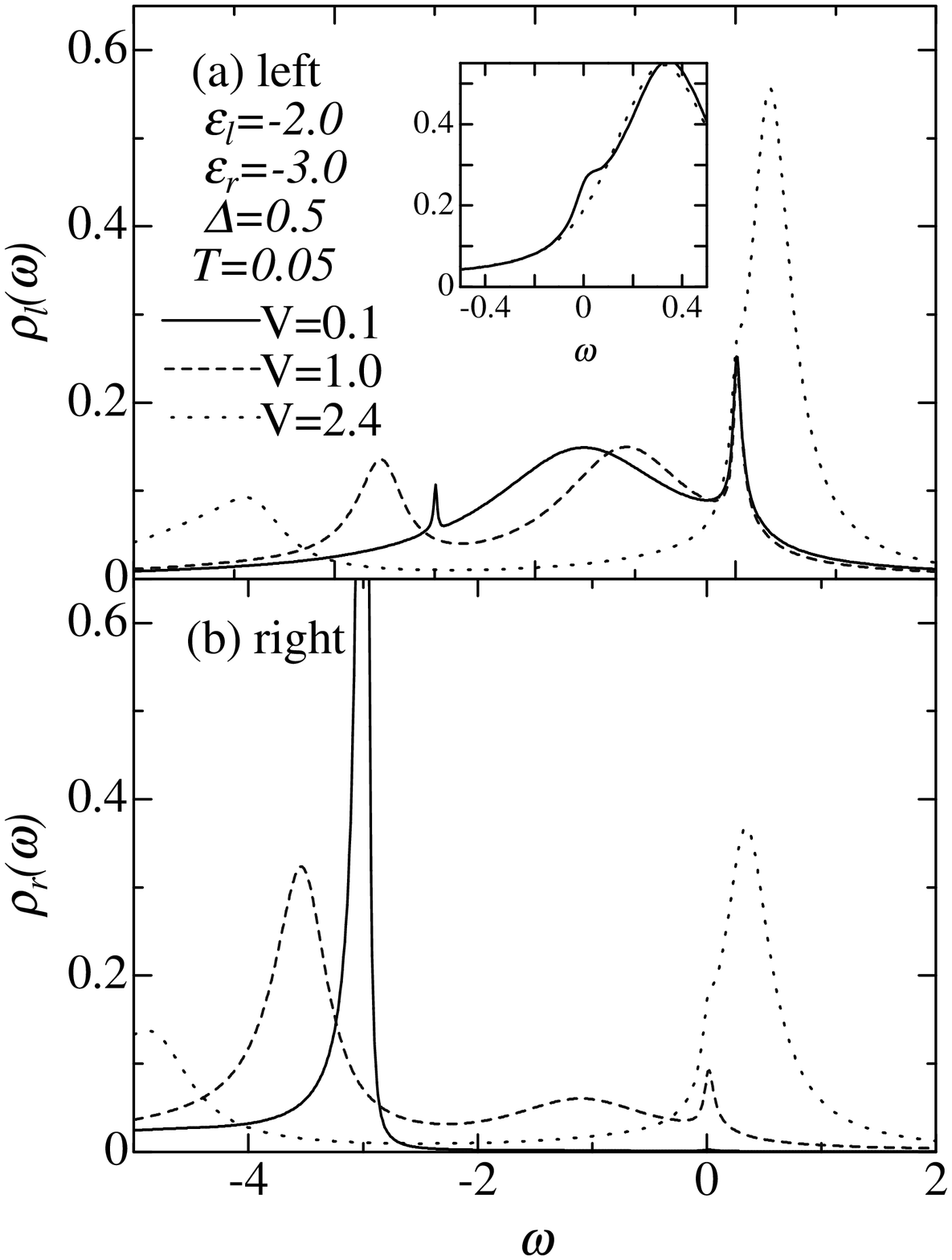}}
\vspace{-20mm}
\caption{Local density of states
in the Kondo regime at $T=0.05$:
 (a) left dot, (b) right dot. 
 The solid, dashed  and dotted lines
correspond to the results for $V=0.1$, $V=1.0$ and $V=2.4$ respectively.
The inset in (a) shows $\rho_{l}(\omega)$ for $V=2.4$, where the 
solid line is for $T=0.05$ and the dotted line for $T=0.2$. 
The effective Kondo temperature is about 0.01 for $V=0$.
The Fermi level is taken to be $\omega =0$. 
}
\label{fig:KondoDOS}
\end{figure}

We start with the Kondo regime shown in  Fig. \ref{fig:Gvdepend}(a).
It is seen that 
the conductance is gradually enhanced with the decrease of the 
temperature 
almost in the whole range of $V$ shown in the figure.
For small $V$, the enhancement is clearly 
attributed to the Kondo effect due to the left dot, which is
indeed confirmed  from the DOS shown in Fig. \ref{fig:KondoDOS}. 
Namely, for small inter-dot coupling ($V=0.1$), it is seen that 
the left dot has the typical DOS
 expected for a single-dot Kondo system, while 
the right dot has a sharp delta-function type
spectrum similarly to an isolated dot. 
When the inter-dot coupling becomes large,
the left- and right-dot levels are mixed up
considerably, leading to 
 the upper and lower levels given by 
\begin{eqnarray}
\epsilon_{\pm}=\frac{1}{2}\left[
(\epsilon_{l\sigma}+\epsilon_{\epsilon_{r\sigma}})
\pm \sqrt{(\epsilon_{l\sigma}-
\epsilon_{\epsilon_{r\sigma}})+4V^{2}}\right].
\end{eqnarray}
 The effective upper level is pushed upward 
 with the increase of $V$, resulting in the increase of
the conductance. It should be noted that 
 the conductance strongly depends on the temperature
even when the effective upper
level is located around the Fermi level.
Its temperature dependence is not due to the Kondo effect, but
is mainly caused by that of the Coulomb peak
itself, as  seen in the corresponding  DOS for the
intermediate inter-dot coupling $V=1.0$ in Fig. \ref{fig:KondoDOS}.
This implies that the 
electron correlations play a crucial role for the 
 conductance even in this parameter regime.
As the inter-dot tunneling further increases, 
the upper level goes beyond the Fermi energy. 
 It is remarkable that 
a tiny resonance peak  again appears around the Fermi level.
 This peak is identified with
the Kondo resonance because it gradually develops with the decrease of 
the temperature, as seen in the inset of  Fig. \ref{fig:KondoDOS}(a).
However, the resulting Kondo temperature 
is rather  small, and  the conductance via the 
Kondo resonance is  hindered by that
caused by the upper level. Therefore in the parameter regime 
$V \sim 2.4$, the conductance is not so much enhanced with the decrease of 
the temperature, although the small Kondo peak indeed exists.

As mentioned before, our NCA results become less reliable 
at low temperatures. To overcome this problem, we 
calculate the conductance  by the SBMF
theory at zero temperature. The results are shown 
in the inset of  Fig. \ref{fig:Gvdepend} (a). 
It is seen that for small $V$ ($\sim 0.1$), the conductance is 
increased almost up to the unitary limit, while 
in the intermediate regime $V \sim 2$, the conductance is 
not so much enhanced.  In the strong $V$ regime, the conductance is 
enhanced, but the effective Kondo peak is 
shifted slightly away from the Fermi level, so it does not 
reach the unitary limit.  In this way, the SBMF 
results confirm that the  conductance calculated by the
NCA approach indeed gives reliable results.

\begin{figure}[h]
\epsfxsize=9cm
\centerline{\epsfbox{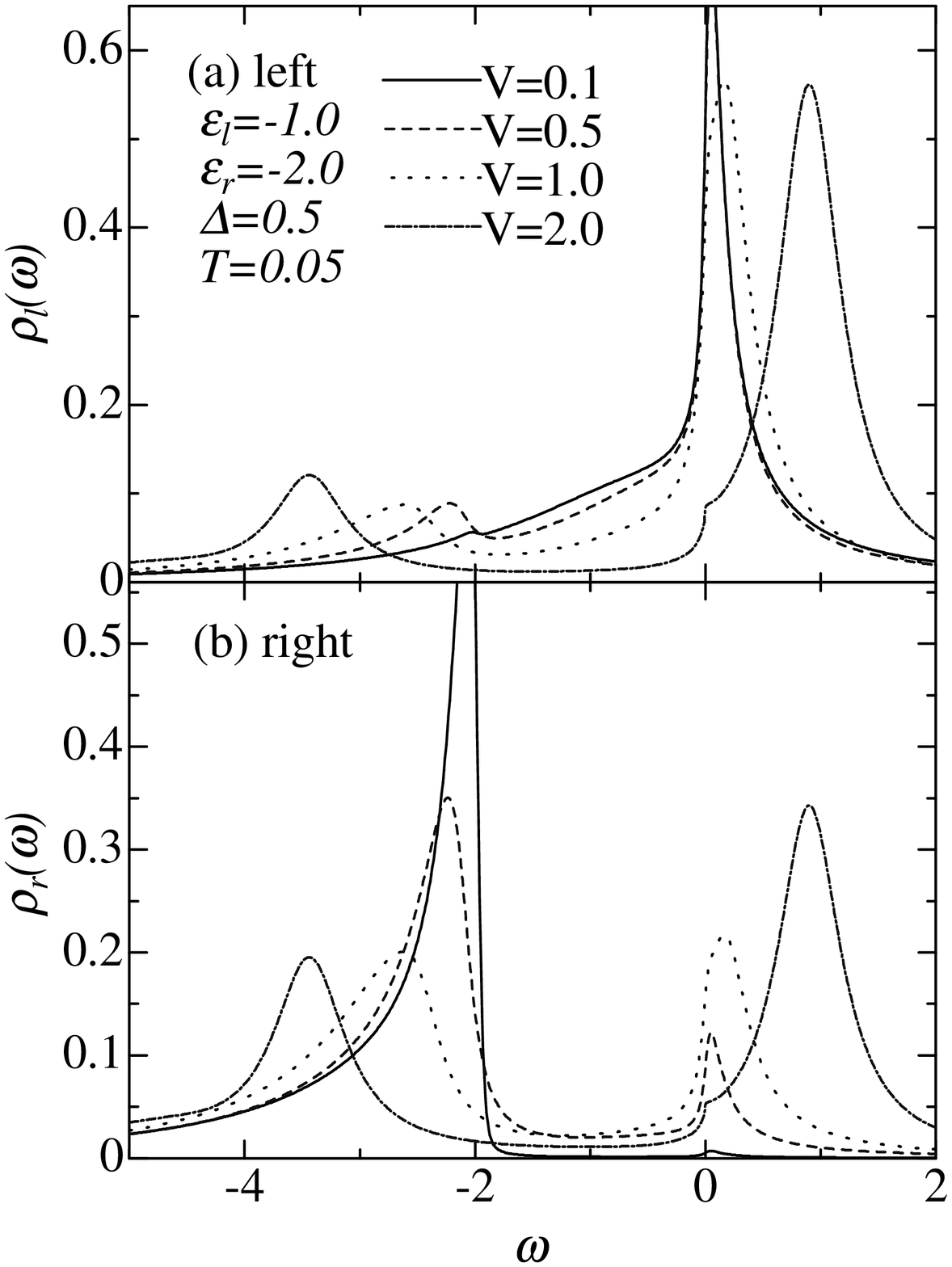}}
\vspace{-20mm}
\caption{Local density of states in the valence-fluctuation
 regime at $T=0.05$:
(a) left dot, (b) right dot.
The solid, dashed,
dotted and  dot-dashed lines 
correspond to  $V=0.1$, $V=0.5$, $V=1.0$ and $V=2.0$ respectively.
  The Fermi level is taken to be $\omega =0$.  
}
\label{fig:valenceDOS}
\end{figure}

We next discuss
 the valence-fluctuation regime.
The calculated  conductance is shown in Fig. \ref{fig:Gvdepend} (b),
and the corresponding  DOS is shown in Fig. \ref{fig:valenceDOS}.
As mentioned above, when the inter-dot tunneling $V$ increases,
two dot levels are separated and thus the upper (lower)
level is shifted  beyond (below) the Fermi level.
Therefore, the conductance is decreased since the Coulomb peak disappears
around the Fermi level.  This is indeed seen in the finite-temperature
NCA results as well as the zero-temperature SBMF results in 
Fig. \ref{fig:Gvdepend}(b).
As mentioned in the case of the Kondo regime, in the large 
$V$ regime, a tiny Kondo peak appears, which can be also seen  in the 
DOS in Fig. \ref{fig:valenceDOS}.
 
\subsection{Tuning the dot-levels}

We now investigate how the conductance is changed when 
the energy level of the left or right dot is 
tuned. This may be experimentally studied 
by changing the gate voltage.

\subsubsection{Conductance control by the left dot}

\begin{figure}[h]
\epsfxsize=9cm
\centerline{\epsfbox{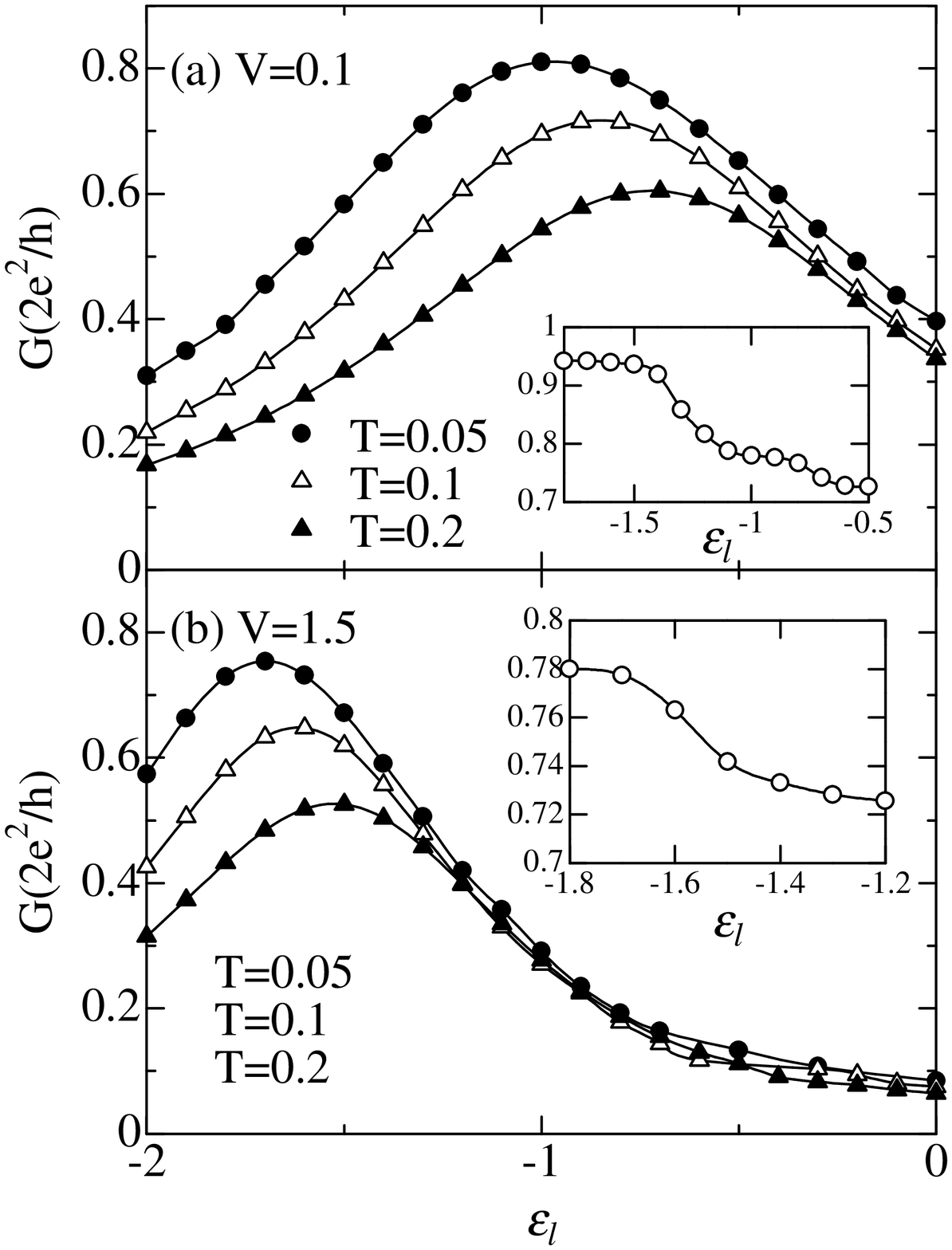}}
\vspace{-20mm}
\caption{Conductance for the T-shaped double-dots system as a 
function of the left-dot level $\epsilon_l$ with 
$\epsilon_r=-3.0$ being fixed: (a) $V=0.1$ and (b) $V=1.5$.}
\label{fig:Geldepend}
\end{figure}

We start by examining  the conductance as a 
function of the left-dot level, $\epsilon_{l}$.
 The results 
are shown in Fig. \ref{fig:Geldepend}. 
Most of the characteristic properties in this case
can be understood from the discussions given  in the 
previous subsection.
In the weak inter-dot coupling regime shown 
in Fig. \ref{fig:Geldepend} (a), 
the conductance is enhanced due to the formation of 
the Kondo resonance as the temperature decreases
when $\epsilon_l$ is sufficiently deep.
As $\epsilon_{l}$ approaches the Fermi level, 
 the conductance is dominated by the current 
via the bare resonance formed around  the
effective energy level of the left dot.
This type of the interpretation is also valid for 
the strong inter-dot coupling regime 
shown in  Fig. \ref{fig:Geldepend} (b), 
for which  the effective level should be replaced by  the  
resultant upper level, $\epsilon_+$.
In this case, the temperature dependence of the conductance 
is understood 
by the same  reason mentioned  in Fig. \ref{fig:Gvdepend} (b). 
We note  that the Kondo peak appears again 
 even for $\epsilon_{l} \sim 0$, but
 the conductance is mainly dominated by 
 the resonance of the upper
 level existing near the Kondo peak.

\subsubsection{Conductance control by the right dot}

\begin{figure}[h]
\epsfxsize=9cm
\centerline{\epsfbox{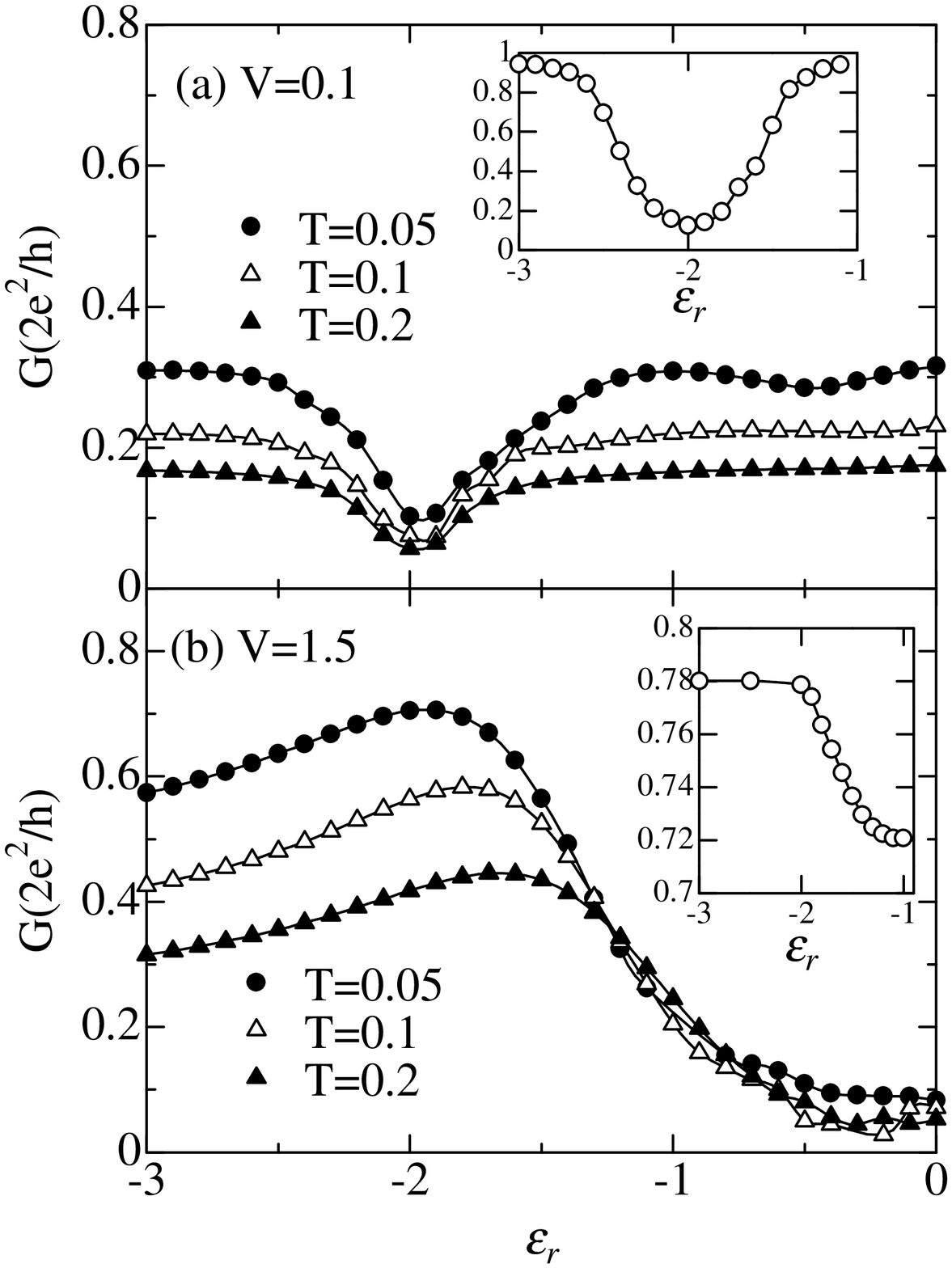}}
\vspace{-20mm}
\caption{Conductance for the T-shaped double-dots system as a 
function of the right-dot level  $\epsilon_r$ 
with $\epsilon_l=-2.0$ being fixed: (a) $V=0.1$ and (b) $V=1.5$.}
\label{fig:Gerdepend}
\end{figure}

We finally  discuss the conductance 
by tuning the energy level of the right dot, $\epsilon_r$.
We  again investigate two cases with small and large inter-dot 
coupling $V$. The results are summarized  in Fig.
\ref{fig:Gerdepend}.
 Let us start with the case of small $V$.
When $\epsilon_r$ of the right dot is deep below 
$\epsilon_l$, the conductance is mainly controlled by the 
Kondo mechanism of the left dot, as already mentioned 
before.  This is also valid for the 
case when $\epsilon_r$ is higher beyond $\epsilon_l$.
These effects are indeed seen in the conductance of Fig.
\ref{fig:Gerdepend}(a). Namely we can see the Kondo enhancement of 
conductance with the decrease of the temperature.

It should be remarked that we encounter a novel interference 
phenomenon, i.e. nontrivial 
suppression of the conductance,  when 
 two bare dot-levels coincide with each other.  This is observed both in
 the NCA results as well as the SBMF results
 in Fig. \ref{fig:Gerdepend}(a) around $\epsilon_r \sim -2.0$.
This may be regarded as a Fano-like effect caused by the 
interference between
two distinct channels,\cite{fano,bogdan,hofstetter2}
 i.e. the direct Kondo resonance
of the left dot and the indirect resonance 
via both of the right and left dots.
As $V$ becomes large, such interference is smeared, and 
cannot be observed in Fig. \ref{fig:Gerdepend}(b).
We wish to mention that this type of dip structure 
can be also seen in Fig. \ref{fig:Geldepend}(a)  if 
 the left level is varied  through the point 
$\epsilon_l \sim \epsilon_r = -3.0$.

\begin{figure}[h]
\epsfxsize=9cm
\centerline{\epsfbox{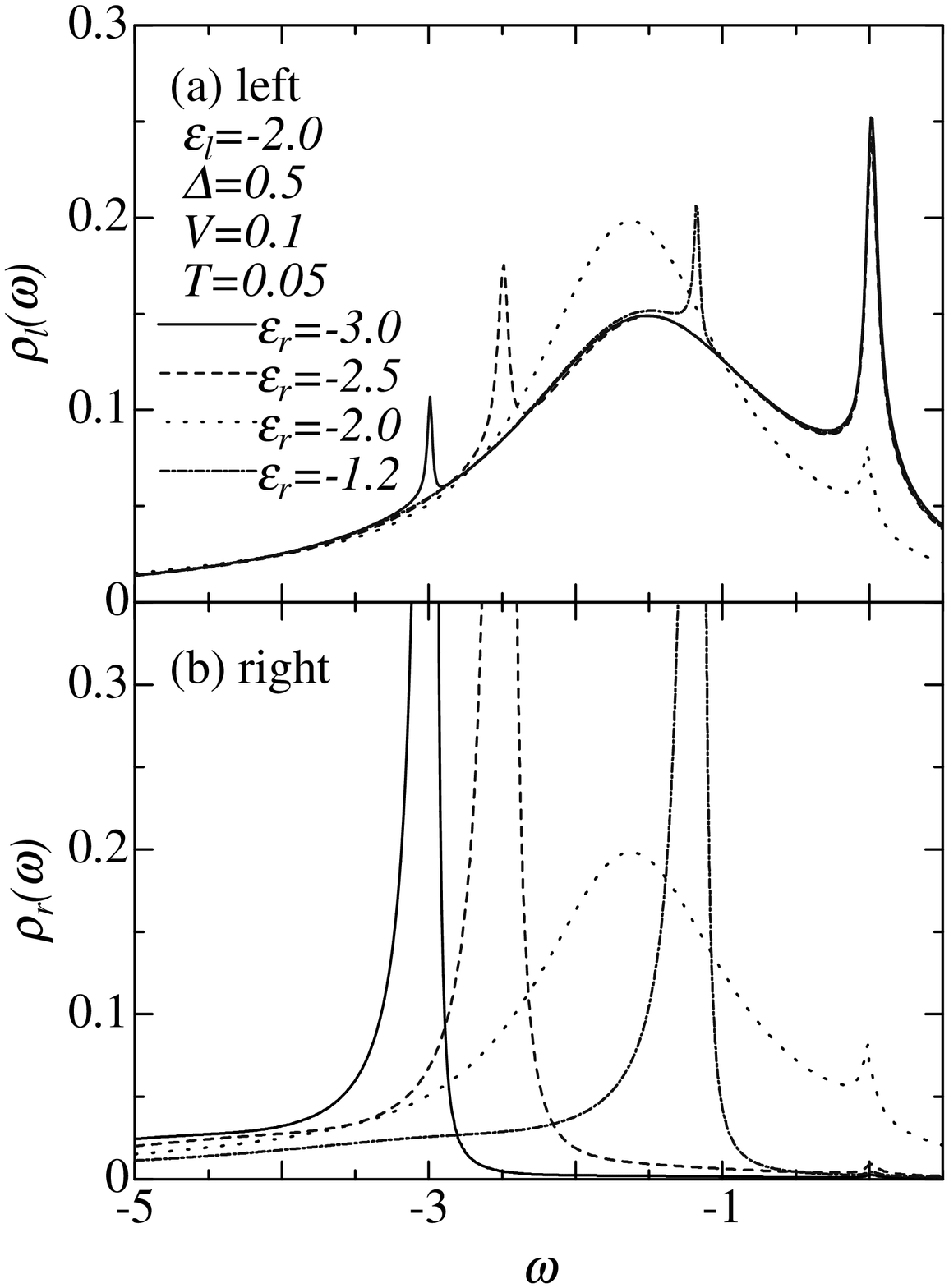}}
\vspace{-20mm}
\caption{Local density of states for $V=0.1$.
The energy level $\epsilon_r$ of the right dot is systematically changed
with a  }
\label{fig:erv01DOS}
\end{figure}

In order to clarify the origin of the interference effect, we 
observe how the DOS changes its character 
when  the right dot level is changed with  $V=0.1$ being fixed.
It is seen in Fig.\ref{fig:erv01DOS}
 that when two localized levels are separated
energetically, the DOS of the left dot is analogous to that for
the ordinary single-dot Anderson model, which  
is slightly modified around $\epsilon_r$ by the
hybridization effect
of the right dot. Therefore, in this parameter regime, the 
conductance is mainly controlled  by the Kondo resonance
of the left dot.  However, when two levels are almost
 degenerate ($\epsilon_r=\epsilon_l=-2.0$),
the left and right states are equally mixed 
together to lift  the degeneracy, and make two copies of
similar DOS. As a consequence, this change around the bare 
levels indirectly modifies the Kondo resonance, namely,
it supresses the formation of the Kondo peak
around the Fermi level, as seen from 
Fig.\ref{fig:erv01DOS} for $\epsilon_r=-2.0$.
This thus gives rise to the decreace in the
conductance. It is to be noted  here that 
such an interference effect occurs when the 
right level is changed in the energy range of the 
order of $\Delta$, while the relevant temperature range,
in which such an effect may be observable, is 
given by the effective Kondo temperature.
Therefore, both of the  bare and 
renormalized (Kondo) energy scales play a crucial role
 in this interference phenomenon.

It is instructive to point out that the interplay of the Kondo 
effect and the interference  effect appears for smaller
$V$, which may be suitable for this effect
to be observed in experiments.

\section{ Summary}

We have studied electron transport in the T-shaped double-dots 
system as a function of the inter-dot coupling as well as the dot levels. 
We have calculated the tunneling conductance by exploiting  the NCA 
at finite temperatures as well as the 
SBMF treatment at zero temperature. It has been found that
 though the electron current flows only through
the left dot, its behavior is considerably affected by 
 the additional right dot via the interference effect. 
 This demonstrates that the 
transport properties of the system
 may be controlled by properly tuning the parameters of
 the right dot which is indirectly
connected to the leads.

In particular, we have found a remarkable interference  phenomenon 
for a small inter-dot coupling in the Kondo regime;
the conductance is suppressed  when two  bare dot-levels 
coincide with each other  energetically, which is
caused  by  the interference effect between two distinct channels
mediated by the left and right dots. 
This effect may be observed 
experimentally around or below the Kondo temperature
by appropriately tuning the gate voltage for the dots in the 
range of the resonance energy due to the inter-dot coupling.
In real double-dots systems, the inter-dot tunneling may be
rather small, which may provide suitable conditions
for our proposed phenomenon to be observed.

\section*{Acknowledgements}
The work is partly supported by a Grant-in-Aid from the Ministry of 
Education, Science, Sports, and Culture.

\newpage


\begin{thebibliography}{99}
\bibitem{glazman}
L.I. Glazman and M.E. Raikh: JETP Lett: {\bf 47} (1988) 452

\bibitem{meir1}
Y. Meir, N.S. Wingreen and P. A. Lee: Phys. Rev. Lett. {\bf 66} (1991) 3048

\bibitem{wingreen}
N.S. Wingreen and Y. Meir: Phys. Rev. B {\bf 49} (1994) 11040

\bibitem{kawabata}
A. Kawabata: J. Phys Soc. Jpn {\bf 60} (1991) 3222

\bibitem{oguri}
A. Oguri, H. Ishii and T. Saso: Phys. Rev. B {\bf 51} (1995) 4715

\bibitem{gold}
D. Goldhaber-Gordon, J. G$\ddot{{\rm o}}$res, M. A. Kastner, 
Hadas Shtrikman, D. Mahalu, and U. Meirav: 
Phys. Rev. Lett. {\bf 81} (1998) 5225

\bibitem{cronenwwett}
S.M. Cronenwett, T.H. Oosterkamp and L.P. Kouwenhoven: 
Science {\bf 281} (1998)  540

\bibitem{schmid}
J. Schmid, J. Weis, Keberl and K.v.Klizing: Physica B {\bf 256} (1998) 182

\bibitem{simmel}
F. Simmel, R.H. Blick, J.P. Kotthaus, W. Wegscheider 
and M. Bicher: Phys. Rev. Lett. {\bf 83} (1999) 804

\bibitem{tarucha}
S. Tarucha, D.G. Austing, T. Honda, R.J. van der Hage 
and L.P. Kouwenhoven: Phys. Rev. Lett. {\bf 77} (1996) 3613

\bibitem{austing}
D.G. Austing, T. Honda and S. Tarucha: 
Jpn. J. Appl. Phys., Part 1 {\bf 36} (1997) 1667

\bibitem{waugh}
F.R. Waugh, M.J Mar, R.M. Westervelt, K.L. Campman and A.C. Gossard: 
Phys. Rev. Lett. {\bf 75} (1995) 705

\bibitem{blick}
R.H. Blick, D. Pfannkuchke, R.J. Haug, K.v. Klitzing and K. Eberl:
 Phys. Rev. Lett. {\bf 80} (1998) 4032

\bibitem{Oosterkamp}
T.H. Oosterkamp, T. Fujisawa, W.G. van der Wiel, 
K. Ishibashi, R.V. Hijman, S. Tarucha and
 L.P. Kouwenhoven: Nature (London). {\bf 395} (1998) 873

\bibitem{aono1}
T.Aono, M. Eto and K. Kawamura: J. Phys. Soc. Jpn {\bf 67} (1998) 1860

\bibitem{aono2}
T. Aono and M. Eto: Phys. Rev. B {\bf 63} (2001) 125327

\bibitem{sakai1}
O. Sakai, Y.Shimizu and T. Kasuya: Solid State Commun. {\bf 75} (1990) 81

\bibitem{izumida}
W. Izumida and O. Sakai: Phys. Rev. B {\bf 62} (2000) 10260

\bibitem{hofstetter}
W. Hofstetter and H. Scheller: Phys. Rev. Lett. {\bf 88} (2002) 016803

\bibitem{vojta}
M. Vojta, R. Bulla and W. Hofstetter: cond-mat/0106458

\bibitem{fano}
U. Fano: Phys. Rev. {\bf 124} (1961) 1866

\bibitem{bogdan}
B. R. Bulka and P. Stefa$\acute{{\rm n}}$ski: Phys. Rev. Lett. {\bf 86} (2001) 5128

\bibitem{hofstetter2}
W. Hofstetter J. K$\ddot{{\rm o}}$nig and H. Scheller: 
Phys. Rev. Lett. {\bf 87} (2001) 156803

\bibitem{Gores}
J. G$\ddot{o}$res, D. Goldhaber-Gordon, S. Heemeyer, and M. A. Kastner: Phys. Rev. B {\bf 62} (2000) 2188 

\bibitem{coleman}
P. Coleman: Phys. Rev. B {\bf 29} (1983) 3035

\bibitem{kuramoto}
Y. Kuramoto:
Z. Phys. B {\bf 53} (1983) 37

\bibitem{bickers}
N. E. Bickers: Rev. Mod. Phys. {\bf 59} (1987) 845

\bibitem{puruschke}
Th. Pruschke and N. Grewe: Z. Phys. B {\bf 74} (1989) 439


\bibitem{imai}
Y. Imai and N. Kawakami: J. Phys. Soc. Jpn {\bf 70} (2001) 1851



\bibitem{jones}
B.A. Jones, G. Kotoliar and A.J. Millis: Phys. Rev. B {\bf 39} (1989) 3415





\bibitem{hewson}
A.C. Hewson, {\it The Kondo Problem to Heavy Fermions} (Cambridge University Press, Cambridge) 1993

\bibitem{meir2}
Y.Meir, N.S. Wingreen and P.A. Lee, Phys. Rev.Lett. {\bf 70} (1993) 2601
\end{thebibliography}
\end{document}